\documentclass[sts]{imsart}

\RequirePackage{amsthm,amsmath,amsfonts,amssymb}
\RequirePackage[numbers,sort&compress]{natbib}
\RequirePackage[colorlinks,citecolor=blue,urlcolor=blue]{hyperref}
\RequirePackage{graphicx}

\startlocaldefs
\theoremstyle{plain}

\newtheorem{prop}{Proposition}[section]

\theoremstyle{definition}

\newcommand{\zb}{\mathbf{z}}
\newcommand{\thetab}{\boldsymbol{\theta}}
\DeclareMathOperator*{\argmin}{arg\,min}
\usepackage{booktabs}
\endlocaldefs

\usepackage{bm}
\usepackage[dvipsnames]{xcolor}

\newcommand{\ech}{\rm\color{black}}

\renewcommand{\ll}{^{(\ell)}}
\newcommand{\ZZ}{\mathcal{Z}}
\newcommand{\lamb}{\bm{\lambda}}

\newcommand{\KL}{\text{KL}}
\newcommand{\qs}{q^\star}

\usepackage[normalem]{ulem}

\begin{document}

\begin{frontmatter}
\title{Vertical Consensus Inference for High-Dimensional Random Partition}
\runtitle{Vertical Consensus Inference for High-Dimensional Random Partition}

\begin{aug}

\author[A]{\fnms{Khai}~\snm{Nguyen}\ead[label=e1]{khainb@utexas.edu}},
\author[B]{\fnms{Yang}~\snm{Ni}\ead[label=e2]{yang.ni@austin.utexas.edu}}
\and
\author[C]{\fnms{Peter}~\snm{Mueller}\ead[label=e3]{pmueller@math.utexas.edu}}


\address[A]{Khai Nguyen is Ph.D. Candidate, Department of Statistics and Data Sciences,
University of Texas at Austin, Texas, USA\printead[presep={\ }]{e1}.}

\address[B]{Yang Ni is Associate Professor, Department of Statistics and Data Sciences,
University of Texas at Austin, Texas, USA\printead[presep={\ }]{e2}.}

\address[C]{Peter Mueller is  Professor, Department of Mathematics,
University of Texas at Austin, Texas, USA\printead[presep={\ }]{e3}.}

\end{aug}

\begin{abstract}
   We review 
   recently proposed  Bayesian 
  approaches for clustering
  high-dimensional data.
  After identifying the main 
  limitations of   available approaches, 
  we  introduce  an alternative framework based  on 
  \textit{vertical consensus 
    inference} (VCI) to mitigate the curse of dimensionality in
  high-dimensional Bayesian clustering.  VCI builds on the idea of
 consensus Monte Carlo by dividing the data into multiple shards
  (smaller subsets of variables), performing posterior inference on each
  shard, and then combining the shard-level posteriors to obtain a
  consensus posterior. The key distinction is that VCI splits the data
  \textit{vertically}, producing \textit{vertical shards} that retain
  the same number of observations but have lower dimensionality.
   We use an entropic regularized Wasserstein barycenter to define
  a consensus posterior. The shard-specific barycenter weights are
  constructed to favor shards that provide meaningful partitions,
  distinct from a trivial single cluster or all singleton clusters,
  favoring balanced cluster sizes and precise shard-specific posterior
  random partitions. We show that 
  VCI can be interpreted as a variational approximation
  to the posterior under a hierarchical model with a generalized
  Bayes prior.
  For relatively low-dimensional problems,
  experiments suggest that
  VCI closely approximates inference based on clustering the entire
  multivariate data.
   For high-dimensional data and in the presence of many
  noninformative dimensions,
  VCI introduces a new framework for model-based and principled inference on
  random partitions. 
  Although our focus here is on random partitions, VCI can be applied
  to any dimension-independent parameters and serves as a bridge to
  emerging areas in statistics such as consensus Monte Carlo, optimal
  transport, variational inference, and generalized Bayes.
\end{abstract}
\ech


\end{frontmatter}

\color{black}
\section{Introduction}
\label{sec:introduction}

\subsection{ High-dimensional clustering}

 Model-based inference for random partitions is often based on
setting up discrete mixture models. Introducing latent indicators that
link data points with the terms in the mixture model defines a random
partition by defining clusters as all items with matching latent
indicator. Setting up a prior model for a random partition then
reduces to introducing a prior probability model for the mixing
measure of the random partition. In fact, it can be shown that subject
to some reasonable symmetry constraints (exchangeability) all random
partitions can be represented this way
\citep{kingman1978representation}. 
Priors placed on
these random probability measures  are known as  Bayesian
nonparametric models, including Dirichlet process
mixtures~\citep{lo1984class,lau2007bayesian}, Pitman–Yor process
mixtures~\citep{pitman1997two}, and other related constructions that
induce distributions on partitions and allow the complexity of the
clustering structure to grow with the data
~\citep{lijoi2005hierarchical,lijoi2007controlling}. We refer
the reader to De Blasi et al. (2013)~\citep{de2013gibbs} for a more general discussion of
Bayesian nonparametric priors on random partitions.
 For low to moderate dimensional data, such inference is well
established and widely used. 

In high-dimensional settings where the number of dimensions $p$ far
exceeds the number of observations $n$, Bayesian clustering methods
based on mixture models can fail in a fundamental way: as the
dimension grows, the posterior distribution over partitions tends to
degenerate to trivial solutions, assigning either all observations to
a single cluster or each observation to its own singleton cluster,
regardless of the true underlying
structure~\citep{chandra2023escaping}. This phenomenon arises from the
interplay between high dimensionality, the choice of likelihood (e.g.,
multivariate Gaussian kernels), and prior specifications, which can
either overly penalize or insufficiently penalize model
complexity~\citep{chandra2023escaping}.

 Beyond model-based clustering based on 
mixture models, high-dimensional clustering is well-known as a
challenging problem. 
One approach is {\em subspace
  clustering}~\citep{parsons2004subspace,agrawal1998automatic}, which
seeks clusters within different combinations of dimensions (i.e.,
subspaces) and, unlike many other methods, does not assume that all
clusters exist in the same set of dimensions. However, these methods are typically designed for point estimation and algorithmic recovery of cluster structure, rather than full probabilistic inference on partitions. Another approach is
{\em projection-based clustering}~\citep{thrun2021using}, which first
reduces the data to a lower-dimensional space before performing
clustering. While often effective computationally, projection can obscure uncertainty and may discard cluster-relevant information when the low-dimensional representation is not well aligned with the latent partition structure. {\em Bagging-based methods}~\citep{dudoit2003bagging} generate
clusters by repeatedly sampling random subsets of the data, clustering
each subset, and then aggregating the results to produce a
dissimilarity measure for the full dataset. Although such methods can improve robustness, they generally provide heuristic aggregation rather than a coherent posterior distribution over partitions.

In the context of
high-dimensional, small-sample data,  Rahman, Johnson and Rao~\citep{rahman2022hyperparameter,rahman2022using}  leverages a transformed Gram matrix to concentrate
{\em feature representations in a low-dimensional space}, enabling accurate
recovery of both cluster assignments and the unknown number of
clusters. \ech In \citep{rahman2022using} they \ech develop a computationally
efficient method based on a transformed left Gram matrix that
preserves cluster structure without requiring dimension reduction or
tuning parameters. Earlier work by Rahman and Johnson (2018)~\citep{rahman2018fast} introduces a
model-based, hyperparameter-free clustering algorithm that exploits
the Gram matrix representation to shift computational dependence from
the feature dimension to the sample size, achieving improved accuracy
and efficiency in high-dimensional
settings. Chandra, Canale and Dunson (2023)~\citep{chandra2023escaping} recently proposed a {\em Bayesian
latent factor mixture model}, which assumes that high-dimensional
observations arise from a lower-dimensional set of latent variables
following a flexible mixture distribution, allowing parsimonious and
accurate modeling of clusters in high-dimensional data. This provides an elegant model-based solution to high-dimensional clustering, although it is tailored to a specific latent variable formulation rather than a general posterior combination framework over subsets of dimensions.

\subsection{Vertical consensus inference}
In this note, we offer a new perspective of the high-dimensional clustering problem that we argue to address some of the mentioned limitations.
Our approach, named vertical
consensus inference (VCI), begins by dividing the data matrix
($n\times p$) into $K>1$ matrices ($n\times p_k$)
 with $\sum_{k=1}^K p_k=p$, called \emph{vertical} shards. We then perform parallel
inference on the random partition within each shard. The final
inference on the random partition is obtained by combining the shard
posteriors using an entropic-regularized Wasserstein
barycenter~\citep{agueh2011barycenters,cuturi2013sinkhorn}, with an
appropriate ground metric defined on the space of partitions. The approach is justified by recognizing it as approximation of the
posterior of a hierarchical model with a  generalized
Bayes~\citep{bissiri2016general} construction via variational
inference~\citep{blei2017variational}. The approach enjoys the
computational advantages of consensus inference or consensus Monte
Carlo (CMC)~\citep{scott2022bayes} while mitigating the curse of
dimensionality in high-dimensional clustering, and it remains
interpretable through a principled modeling and inference
framework. Experiments on real datasets demonstrate that
 in low dimensional problems 
VCI closely
approximates  inference based on clustering the entire
multivariate data, 
while  it performs favorably  in settings with many noninformative
dimensions and in high-dimensional regimes,  as they arise, for
example, in inference for  single-cell
data. While we focus on random partitions, VCI is applicable  for any
 parameters of interest that are dimension-independent. VCI bridges four emerging areas:
consensus Monte Carlo, optimal transport, variational inference, and
generalized Bayes.

VCI leverages the
divide-and-conquer principle of CMC~\citep{scott2022bayes}. The key
distinction is that VCI 
partitions data into vertical shards  (splitting outcome
dimensions),  rather than the horizontal
shards  (splitting the sample size) 
used in conventional CMC. Traditional CMC is primarily
motivated by the poor scalability of Bayesian inference algorithms,
especially sampling-based methods, with respect to the number of
observations $n$. In contrast, VCI  is designed to address 
the curse of dimensionality in random partition models and, more
broadly, in models with dimension-independent  parameters. 

The concept of vertical partitioning has been explored in federated learning, particularly in vertical federated learning (VFL)~\citep{liu2024vertical}, positioning VCI as a VFL-like framework for Bayesian random partition models. Related work in distributed clustering~\citep{strehl2002cluster} investigates clustering from partial sets of dimensions, but only at the point-estimate level, whereas VCI operates directly at the posterior level. Furthermore, VCI can be interpreted as an instance of subspace clustering~\citep{parsons2004subspace} within the Bayesian random partition context. Finally, VCI can also be interpreted as an approximation of model-based inference in high-dimensional Bayesian random partition models by way of the already mentioned generalized Bayes construction.

Like CMC, VCI
 naturally lends itself to 
distributed and parallel computing: each shard can be assigned to a
separate worker machine~\citep{huang2005sampling}, which independently
performs posterior inference using methods such as Markov chain Monte
Carlo (MCMC)~\citep{geyer1992practical}, sequential Monte Carlo
(SMC)~\citep{doucet2001introduction}, variational inference
(VI)~\citep{blei2017variational},  expectation propagation
(EP)~\citep{minka2001expectation}  or  approximate Bayesian
computation~\citep{beaumont2002approximate}. These methods
can be used either to draw samples from the posterior or to construct
a tractable posterior approximation.

 For the choice of the consensus mechanism in general CMC, 
there are many available methods to combine posterior beliefs across all
shards. 
Early work on consensus Monte Carlo distributes data across machines
and combines local posterior draws via weighted
averaging~\citep{huang2005sampling, scott2022bayes}. Other approaches
approximate the full posterior as a product of local posterior
densities using parametric, semiparametric, or nonparametric
estimators~\citep{neiswanger2014asymptotically,
  white2015piecewise}. Alternative strategies include the Weierstrass
sampler~\citep{wang2013parallelizing},   sequential Monte Carlo and
mixtures~\citep{scott2017comparing}, iterative consensus with global
CMC~\citep{rendell2020global}, geometric median aggregation in
reproducing kernel Hilbert spaces~\citep{minsker2014scalable},
Wasserstein
barycenter~\citep{srivastava2015wasp,srivastava2018scalable},
variational aggregation of local
posteriors~\citep{rabinovich2015variational}, and likelihood inflating
sampling (LISA), which inflates the likelihood to make each shard
posterior resemble the full-data
posterior~\citep{entezari2018likelihood}. CMC for random partition
models was introduced in~\citep{ni2020consensus}; however, 
still relying on horizontal splitting, rather than the vertical
splitting employed in VCI.



\section{vertical consensus inference for Random Partition Models}
\label{sec:VCI}

\subsection{Wasserstein Consensus Monte Carlo}
\label{subsec:WASP}
VCI uses a version of  Wasserstein barycenter to define
a consensus posterior. The use of a Wasserstein barycenter for CMC
 was 
first introduced in~\citep{srivastava2015wasp,srivastava2018scalable}.
 We introduce some notation by way of a brief review of Wasserstein
distance. 
Given a metric space $\mathcal{X}$
equipped with a ground metric $c:\mathcal{X} \times \mathcal{X}\to
\mathbb{R}_+$ ($\mathbb{R}_+$ is the positive real line), Wasserstein
distance~\citep{villani2009optimal,peyre2020computational} between  two probability measures $p_1\in \mathcal{P}_c(\mathcal{X})$ and $p_2\in
\mathcal{P}_c(\mathcal{X})$ is defined as follows:
\begin{align}
\label{eq:W}
    &W_c(p_1,p_2) = \min_{\pi \in \Pi(p_1,p_2)} \int_{\mathcal{X}\times \mathcal{X}} c(x_1,x_2) \mathrm{d} \pi(x_1,x_2), 
\end{align}
where $\mathcal{P}_c(\mathcal{X})$ denotes the set of all probability measures on $\mathcal{X}$ such that for any $p \in \mathcal{P}_c(\mathcal{X})$, there exists $x_0 \in \mathcal{X}$ with $\int_{\mathcal{X}} c(x,x_0) \mathrm{d}p(x) < \infty$, and   $\Pi(p_1,p_2)$ denotes the set of couplings, i.e., joint probability measures whose marginals are $p_1$ and $p_2$, respectively. If 
$c$ is a valid metric between two partitions, then the Wasserstein distance is also a valid metric on $\mathcal{P}_c(\mathcal{X})$~\citep{villani2009optimal}.

With Wasserstein distance, we can generalize the notion of
``averaging" to $\mathcal{P}_c(\mathcal{X})$.  The generalized average
is called barycenter, or Fréchet
mean~\citep{veldhuis2002centroid,banerjee2005clustering}. Given $K>1$
probability measures $p_1,p_2,\ldots,p_K \in
\mathcal{P}_c(\mathcal{X})$ and a ground metric $c$ on $\mathcal{X}$,
the Wasserstein barycenter~\citep{agueh2011barycenters} of
$p_1,\ldots,p_k$ is the defined as follows: 
\begin{align} \label{eq:barycenter}
    \bar{p}=\argmin_{p \in \mathcal{P}(\mathcal{X})} \sum_{k=1}^K \lambda_k W_{c}(p,p_k),
\end{align}
where $\boldsymbol{\lambda}=(\lambda_1,\ldots,\lambda_K) \in \Delta^K$ is a simplex vector of barycenter weights which control the contributions of \textit{marginals}  $p_1,\ldots,p_k$ to the barycenter. Here, we use $\Delta^K$ to denote the $K$-simplex.

Using the uniform weights, Srivastava et al (2015,
2018)~\citep{srivastava2015wasp,srivastava2018scalable}  introduced
 a CMC scheme  using a Wasserstein consensus posterior.
Consider a  dataset of $n>0$ samples
$
X = (X_1, \dots, X_n),
$
where $X_i \in \mathcal{X}^p \text{ for } i = 1, \dots, n,$
 with 
$\mathcal{X}$ denoting the data space (e.g., $\mathbb{R}$) of $p > 0$
dimensions.  In statistical inference, our primary
goal is to infer latent variables $\thetab$ that capture the
underlying structure of the data $X$.
 CMC implements inference for
global parameters  $\thetab$
that are used across shards.
CMC, as well as the upcoming discussion,   assume
that any parameters specific to particular  shards  are integrated out. 
Under a model-based Bayesian
framework, inference on $\thetab$ proceeds via the posterior
distribution 
$
p(\thetab \mid X) = \frac{p(\thetab, X)}{p(X)},
$
derived from the joint distribution $p(\thetab, X)$.  We divide $X$
into $K>1$ shards $X^{(1)},\ldots,X^{(K)}$ with
$X^{(k)}=(X^{(k)}_1,\ldots,X^{(k)}_{n_k})$ ($\sum_{k=1}^K n_k =
n$).
 In~\citep{srivastava2015wasp,srivastava2018scalable},
the shards are defined by splitting the data into subsets of
observations (in contrast to splitting along the dimensions of the
outcome, which we will introduce later). Let $p_k(\thetab\mid X^{(k)})
= \frac{p_k(\thetab, X^{(k)})}{p_k(X^{(k)})}$ is the $k$-th shard
posterior and $c=\|\cdot\|_2^2$, the Wasserstein consensus posterior
is the barycenter of $p_1(\thetab\mid X^{(1)}),\ldots, p_K(\thetab\mid
X^{(K)})$ with the uniform barycenter weights
in~\eqref{eq:barycenter}. 



\subsection{
  A Vertical Consensus Inference Approach for High-Dimensional Clustering}
\label{subsec:proposed_methods}
 In contrast to the horizontal (along samples) split of CMC, the
proposed VCI scheme  splits the \textit{feature} dimension into
$K>1$ \textit{vertical} shards:
$X_i = (X_i^{(1)},\dots,X_i^{(K)})$ for $i=1,\ldots,n$, 
$X_i^{(k)} \in \mathbb{R}^{p_k}$. For simplicity, we assume that $\sum_{k=1}^K p_k = p$ (non-overlapping shards), although the case $\sum_{k=1}^K p_k > p$ is also valid. The
shard splitting process can be performed manually or arise from
distributed data storage or some prior knowledge about data e.g.,
groups of genes.
 In the upcoming discussion, we focus on inference for a random
partition
$\thetab=\zb$, keeping in mind that the approach remains valid for any
other global parameter $\thetab$. Here  
$\zb =(z_1,\ldots,z_n)$ denotes a partition represented by cluster
membership indicators $z_i=h$ if the $i$-th data point in cluster
$h$. For the $k$-th shard $X^{(k)} = (X^{(k)}_1,\ldots,X^{(k)}_n)$, 
 let $p_k = p_k(\zb \mid X^{(k)})$ denote  the shard-specific
posterior (on the partition) for shard $k$. 
We then define a consensus posterior
on $\zb$ as Wasserstein barycenter of shard posteriors $p_1(\zb
\mid X^{(1)}),\ldots,p_k(\zb \mid X^{(K)})$.  

\paragraph*{ Wasserstein barycenter.}
Let $\ZZ$ denote 
the space of partitions of $n$ samples
 represented by cluster membership indicators,  i.e., $\zb \in
 \ZZ$. 
As $\ZZ$ is a finite space, given a ground metric
$c:\ZZ \times \ZZ\to \mathbb{R}_+$, Wasserstein
distance~\eqref{eq:W} between two \textit{discrete} probability measures $p_1\in
\mathcal{P}(\ZZ)$ and $p_2\in \mathcal{P}(\ZZ)$ can be rewritten
 as follows:
\begin{align*}
    &W_c(p_1,p_2) = \min_{\pi \in \Pi(p_1,p_2)} \sum_{i=1}^{|\ZZ|}
      \sum_{j=1}^{|\ZZ|} c(\zb_i,\zb_j) \pi(\zb_i,\zb_j) ,  
\end{align*}
For computational
 reasons (to turn the minimization into a strongly convex
 optimization), in practice
 one often uses 
entropic approximation of Wasserstein distance~\citep{cuturi2013sinkhorn}:
\begin{align}
\label{eq:eOT}
    W_{c,\epsilon}(p_1,p_2)  &= \min_{\pi \in \Pi(p_1,p_2)}\sum_{i=1}^{|\ZZ|} \sum_{j=1}^{|\ZZ|} c(\zb_i,\zb_j) \pi(\zb_i,\zb_j) \\
    &\qquad-\epsilon H_\pi(\zb_1,\zb_2), \nonumber
\end{align}
where $H_\pi(\zb_1,\zb_2) = - \sum_{i=1}^{|\ZZ|} \sum_{j=1}^{|\ZZ|} \pi (\zb_1,\zb_2)\log \pi(\zb_1,\zb_2)$ is the entropy, and $\epsilon>0$ is the regularization strength coefficient. When $p_1$ and $p_2$ have at most $m$ atoms, the time complexity for $W_{c,\epsilon}(p_1,p_2)$ is $\mathcal{O}(m^2)$~\citep{altschuler2017near} compared to  $\mathcal{O}(m^3 \log m)$ of $W_{c}(p_1,p_2)$~\citep{peyre2020computational}. 

We then define a consensus posterior on $\zb$ from shard posteriors
$p_1(\zb \mid X^{(1)}),\ldots,p_k(\zb \mid X^{(K)})$
 using barycenter~\eqref{eq:barycenter} with $W_{c,\epsilon_k}$: 
\begin{align}
    \label{eq:consesus_posterior}
    \bar{p}(\zb\mid X,\boldsymbol{\lambda}) = \argmin_{p \in \mathcal{P}(\ZZ)} \sum_{k=1}^K \lambda_k W_{c,\epsilon_k}(p,p_k),
\end{align}
where $(\lambda_1,\ldots,\lambda_K) \in \Delta^K$ is a simplex vector of barycenter weights. The probability measures $\bar{p}(\zb\mid X,\boldsymbol{\lambda})$ is well-defined and it reflects the belief on $\zb$ from $X$. 

\paragraph*{Ground metric.}
For the choice of ground metric $c$, there are many options, e.g., Binder
loss~\citep{binder1978bayesian},
variation of information (VoI)~\citep{meilua2007comparing}, normalized
variation of information, information distance, normalized
information distance~\citep{vinh2009information},
one minus adjusted Rand index~\citep{rand1971objective,frischimproved2009}, generalized
Binder, and generalized VoI~\citep{dahl2022search}. In this work, we
focus on using VoI distance, which is a valid metric on $\ZZ$.
 Assume then two partitions (in the upcoming construction, 
posterior Monte Carlo samples) of $n$ elements, represented as cluster
membership indicators
$\zb_\ell = (z_{\ell,1},\ldots,z_{\ell,n})$, $\ell=1,2$. 
We define  cluster proportions as 
$p\ll_j = \frac{1}{n}\sum_{i=1}^n \mathbf{1}(z_{\ell,i} = j)$ for $\ell=1,2$ and $j=1,\ldots,\mathcal{H}(\zb_\ell)$ ($\mathcal{H}(\zb_\ell)$ is the number of clusters of $\zb_\ell$), 
and joint proportions
\begin{align*}
p^{(12)}_{jh} = \frac{1}{n}\sum_{i=1}^n \mathbf{1}(z_{1i} = j,\, z_{2i} = h),
\end{align*}
for $j=1,\ldots,\mathcal{H}(\zb_1)$ and $h=1,\ldots,\mathcal{H}(\zb_2)$.
The entropies are
$H(\zb_\ell) = -\sum_{j=1}^{\mathcal{H}(\zb_l)} p\ll_j \log p\ll_j$ for $\ell=1,2$,
and their mutual information is
\begin{align*}
I(\zb_1,\zb_2)
=
\sum_{j=1}^{\mathcal{H}(\zb_1)} \sum_{h=1}^{\mathcal{H}(\zb_2)} p^{(12)}_{jh}\log\frac{p^{(12)}_{jh}}{p^{(1)}_j p^{(2)}_h}.
\end{align*}
The VoI distance between the two partitions $\zb_1$ and $\zb_2$ is defined as
\begin{align*}
\mathrm{VoI}(\zb_1,\zb_2)
=
H(\zb_1) + H(\zb_2) - 2 I(\zb_1,\zb_2).
\end{align*}
We use the VoI distance as the ground metric in~\eqref{eq:eOT}.

\paragraph*{Barycenter weights.}
In~\eqref{eq:consesus_posterior}, the barycenter weight
$\lambda_k$ control the contribution of the $k$-th shard
posterior to the consensus posterior. We can make $\lambda_k$ to be a
function of $p_k$ to prioritize  shard with meaningful partitions. For example, one could use expected entropy to penalize a
single cluster and to prioritize shard posteriors with more balanced clusters: 
\begin{align*}
    &\omega_k^{H}(p_k) = \mathbb{E}[H(\zb) \mid X^{(k)}],\qquad k=1,\ldots,K.\\
    &\lamb^{H}= P_{\Delta^K} (\boldsymbol{\omega}^H),
\end{align*}
where $\boldsymbol{\omega}^H =
(\omega_1^{H}(p_k),\ldots,\omega_K^{H}(p_k))$ and $P_{\Delta^K}:
\mathbb{R}_+^{K} \to \Delta^K$ is the projection function to the
simplex, e.g., we can set $P_{\Delta^K}(\omega_1,\ldots,\omega_K) = \left(\frac{\omega_1^t}{\sum_{k=1}^k \omega_k^t},\ldots,\frac{\omega_K^t}{\sum_{k=1}^k \omega_k^t}\right)$ for any $t\geq 1$, or softmax function. 

Instead of this simple example, we recommend
 more structured 
choices that better reflect investigator
preferences for non-trivial partitions (different from $n$ singletons,
or a single cluster), balanced (with comparable cluster
sizes), and precisely estimated partitions (high entropy of the random
partition). In our implementation, we use the following construction:

\begin{align*}
&\omega_k^P(p_k) 
= \underbrace{\mathbb{E}\left[\frac{4(\tilde H - 1)(n - \tilde H)}{(n - 1)^2} \mid X^{(k)}\right]}_{\text{(I): cluster complexity}}
\\&\underbrace{\mathbb{E}\left[\exp\!\big(-a\,E(\zb) \big) \mid X^{(k)} \right]}_{\text{(II): entropy control}}
\underbrace{\left(1 - 4U\right)}_{\text{(III): uncertainty penalty}}, \nonumber\\
&\lamb^{P}= P_{\Delta^K} (\boldsymbol{\omega}^P),
\end{align*}
where $a \in \mathbb{R}$, and
\begin{align*}
    &\tilde H 
= \exp(-H(\zb)),  
E(\zb) 
= - \frac{H(\zb)}{\log(\mathcal{H}(\zb))}, \nonumber
\\
&
p_{ij} = \mathbb{P}(z_i=z_j\mid X^{(k)}), 
U = \frac{2}{n(n-1)} \sum_{i<j} p_{ij}\big(1 - p_{ij}\big).
\end{align*}
For the first term (I), since $\tilde H \in [1, n]$ corresponding to the entropy range $[0, \log n]$, the product $(\tilde H - 1)(n - \tilde H)$ ranges from $0$ to $(n-1)^2/4$. Consequently, term (I) lies in $[0, 1]$, favoring the number of clusters to avoid extremes of 1 (a single cluster) or $n$ (all singletons). For term (III), $U$ ranges over $[0, 1/4]$, so term (III) also lies in $[0, 1]$, thereby penalizing uncertainty in clustering. For term (II), its range is $[\min\{1, e^{-a}\}, \max\{1, e^{-a}\}]$. When $a < 0$, term (II) favors larger entropy, whereas when $a > 0$, it penalizes entropy. Overall, $\lambda^P$ strongly penalizes both the case of a single cluster and that of all singletons, favoring shards with low posterior  uncertainty, and controls the desired entropy through the parameter $a$.  Other constructions can also be used, depending on the desired properties of shard posteriors. 

Again, we recall that  VCI can extend beyond partitions for any
\textit{dimension-independent} variables equipped with a well-defined
metric on their space of realizations.
 The definition of barycenter weights would have to be adjusted
accordingly.
But 
in this work, we focus on random partitions, as they constitute a
widely studied inference target.

\subsection{Interpretation via Generalized Bayes Hierarchical Model and Variational Inference}
\label{subsec:interpretation}
While the consensus posterior in~\eqref{eq:consesus_posterior} is
well-defined, it
 was not constructed as principled model-based inference.
We now show that $\bar{p}$ can  in fact  be interpreted as
variational approximation of the posterior under the following hierarchical
model. 
\begin{align}
    &p(X^{(k)} \mid \zb^{(k)})  = p_k (X^{(k)} \mid \zb^{(k)}), \label{hier}\\
    &p(\zb^{(k)} \mid \zb)  \propto \exp \left[-\zeta_k
      c(\zb^{(k)},\zb)\right],
      \nonumber\\
    &p(\zb) \propto \prod_{k=1}^K C_k (\zb), \nonumber
\end{align}
where $C_k (\zb) = \sum_{ \zb^{(k)}\in \ZZ} \exp \left[-\zeta_k
  c(\zb^{(k)},\zb)\right] $ ($C_k(\zb) <\infty$ as $\ZZ$ is a
finite space), $\zeta_k>0\, \forall k=1,\ldots,K$ , and $c$ is a ground
metric between partition. The  definition of $p(\zb^{(k)} \mid
\zb)$
in the second line  is analogous to centered partition
processes~\citep{paganin2020centered} and can be
 seen as a generalized Bayes~\citep{bissiri2016general} prior. It 
constructs a conditional distribution by way of a loss function (metric
$c$) rather than starting with an assumed sampling or prior model.

Next, we consider variational
inference~\citep{blei2017variational} to approximate the posterior
on $\zb$ under model \eqref{hier}.
We consider a
variational family $\mathcal{Q} \subset \mathcal{P}(\ZZ^{K+1})$ and
solve for
\begin{align}
\label{eq:VI}
  \qs = \arg \min_{q \in \mathcal{Q}} \KL(q,p),
\end{align}
where $\KL(q,p)$ is the Kullback–Leibler divergence. It is well-known
that~\eqref{eq:VI} admits a dual problem as minimizing a negative evidence
lowerbound (NELBO): 
\begin{align}
\label{eq:ELBO}
    &\min_{q \in \mathcal{Q}} \mathcal{L}(q):=\mathbb{E}_{(\zb,\zb^{(1):(K)})\sim q}[-\log p(\zb,\zb^{(1):(K)},X)] \\
    &\qquad \qquad \qquad - H_q(\zb,\zb^{(1):(K)}). \nonumber
\end{align}

In~\eqref{eq:ELBO}, we minimize the NELBO over a joint distribution on $(\zb,\zb^{(1)},\ldots,\zb^{(K)})$. We can cast the above problem as a nested optimization problem, i.e., maximizing over marginals of the joint distribution and maximizing over the couplings of all marginals~\citep{wu2026extending}. Let $q_0(\zb),q_1(\zb^{(1)}),\ldots,q_K(\zb^{(K)})$ be marginals of $q$, we can rewrite~\eqref{eq:ELBO} into:
\begin{align*}
    \min_{q_0 \in \mathcal{Q}_0,q_1 \in \mathcal{Q}_1,\ldots,q_K \in \mathcal{Q_K}} \min_{q \in \Pi(q_0,q_1,\ldots,q_K)}\mathcal{L}(q),
\end{align*}
where $\mathcal{Q}_0,\mathcal{Q}_1,\ldots,\mathcal{Q}_K \subset \mathcal{P}(\ZZ)$ are corresponding variational families for the marginals, and $\Pi(q_0,q_1,\ldots,q_K)$ is the set of joint probability measures of $(\zb,\zb^{(1)},\ldots,\zb^{(K)})$ with marginals $q_0,q_1,\ldots,q_K$ respectively.
 At this moment we have written the variational problem to include
optimization w.r.t. the coupling $q$. In the following result we go
one step further and relate the variational solution to a Wasserstein
barycenter of $q_1,\ldots,q_K$ (to be further related to the VCI posterior
\eqref{eq:consesus_posterior} below). 

\begin{prop}
For $q_1,\ldots,q_K$  be marginals of the variational posterior $q(\zb^{(1)},\ldots,\zb^{(k)})$, the entropic Wasserstein barycenter~\eqref{eq:consesus_posterior}
  corresponds to an upper bound for the
minimal NELBO over all possible couplings of $q_1,\ldots,q_K$. Specifically:
\begin{align}
&\min_{q \in \Pi(q_0, q_1, \dots, q_K)} \mathcal{L}(q)  \leq \sum_{k=1}^K \zeta_k W_{c, \frac{1}{K \zeta_k}}(q_0, q_k) \label{eq:prop_equation}   \\
&- \sum_{k=1}^K\mathbb{E}_{ \zb^{(k)}\sim q_k}\left[\log  p_k (X^{(k)} \mid \zb^{(k)})\right] +C, \nonumber
\end{align}
where  $C=\sum_{\zb\in\ZZ} \prod_{k=1}^K C_k(\zb) >0$, for $C_k(\zb)$ in~\eqref{hier}, is a constant.
\end{prop}
The proof appears in Appendix A.
 Next, fix $q_k$ as the shard posteriors, 
$q_k(\zb^{(k)}) = p_k(\zb^{(k)} \mid X^{(k)})$, minimizing the right hand side of~\eqref{eq:prop_equation} with respect to $q_0$ leads to $\bar{p}(\zb\mid X,\lamb)$
in~\eqref{eq:consesus_posterior} as $\mathbb{E}_{ \zb^{(k)}\sim
  q_k}\left[\log  p_k (X^{(k)} \mid \zb^{(k)})\right]$ is a constant
with respect to $q_0$.
 In summary, 
solving the barycenter in~\eqref{eq:consesus_posterior}  can be seen as minimizing an upper bound of the NELBO with respect to the variational posterior (consensus posterior) $q_0$. 
 \begin{figure*}[!t]
    \centering
    \includegraphics[width=0.75\linewidth]{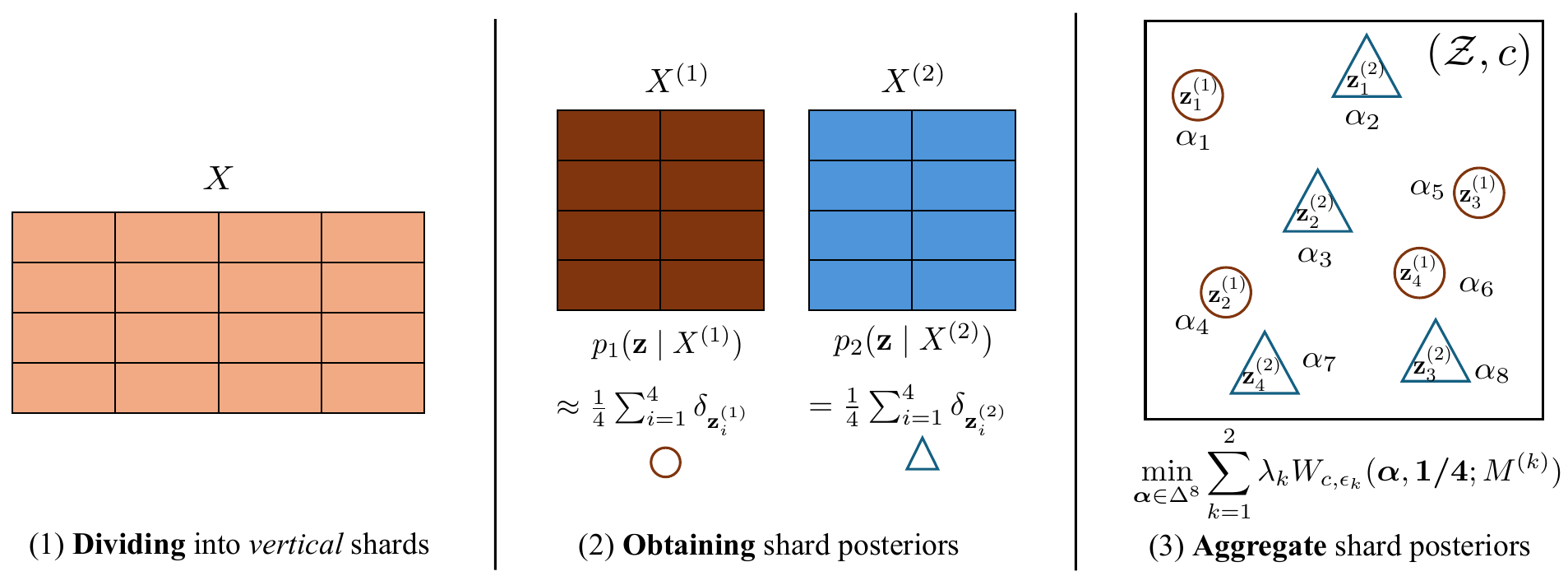}
    \vspace{-1,5em}
    \caption{A simple illustration of VCI with 2 shards and 4 posterior samples per shard.}
    \label{fig:VCI}
     \vspace{-2em}
\end{figure*}
 In~\eqref{eq:prop_equation}, if we divide  $\zeta_k$ by $\sum_{k=1}^K
\zeta_k$, the minimization with respect to $q_0$ stays the same. Therefore, for the
correspondence to~\eqref{eq:consesus_posterior}, we have $\zeta_k =
\frac{1}{K\epsilon_k}$ and  $\lambda_k = \frac{\zeta_k}{\sum_{k=1}^K
  \zeta_k}$. We recall that multiplying
the entropy $H_q(\zb,\zb^{(k)})$
in~\eqref{eq:entropy_inequality} in the proof of the proposition by any real positive number smaller
than 1 still preserves a bound. Hence, the choice of $\lambda_k$
and $\epsilon_k$ is flexible as in~\eqref{eq:consesus_posterior}.

The gap in~\eqref{eq:prop_equation} is tight when the entropy inequality~\eqref{eq:entropy_inequality} and the inequality  in~\eqref{eq:star_inequality} in the proof of the proposition (in Appendix A)  are tight. For~\eqref{eq:entropy_inequality}, the inequality is tight when  $H_q(\zb^{-(k)}\mid \zb,\zb^{(k)})=0$, meaning that the partition (up to relabeling) 
$\zb^{-(k)}$
 are deterministically determined by  $\zb,\zb^{(k)}$ under the variational posterior $q$ for all $k=1,\ldots,K$. Consequently, the uncertainty under $q$ of the $(\zb,\zb^{(1)},\ldots,\zb^{(K)})$ is fully determined once 
$\zb$ and any $\zb^{(k)}$ are known. The inequality~\eqref{eq:star_inequality} is tight when the optimal variational posterior has a star structure, i.e., we can decompose $q(\zb, \zb^{(1)}, \dots, \zb^{(K)}) = q_0(\zb) \prod_{k=1}^K q(\zb^{(k)} \mid \zb)$.

\subsection{Computational Aspects} 
\label{subsec:computation}
We now discuss how to solve the entropic Wasserstein barycenter
problem in practice. As the space of partitions is discrete, we can
write  any  probability measure on $\ZZ$ as
$p(\zb)=\sum_{i=1}^{\ZZ} \alpha_i \delta_{\zb_i}$ where
$\boldsymbol{\alpha}=(\alpha_1,\ldots,\alpha_{|\ZZ|}) \in
\Delta^{|\ZZ|}$. Without  loss of  generality, we can write the $k$-th
shard posterior $p_k(\zb \mid X^{(k)})= \sum_{i=1}^{\ZZ}
\alpha_{i}^{(k)} \delta_{\zb_i}$ where
$\boldsymbol{\alpha}^{(k)}=(\alpha_{1}^{(k)},\ldots,\alpha_{|\ZZ|}^{(k)})
\in \Delta^{|\ZZ|}$. Therefore, we can rewrite the consensus posterior
problem in~\eqref{eq:consesus_posterior} as: 
\begin{align}
\label{eq:fixed_support_barycenter}
    \min_{\boldsymbol{\alpha} \in \Delta^{|\ZZ|} } \sum_{k=1}^K \lambda_k  W_{c,\epsilon_k}(\boldsymbol{\alpha},\boldsymbol{\alpha}^{(k)};M),
\end{align}
where $M \in \mathbb{R}_+^{|\ZZ| \times |\ZZ|}$ is the cost matrix with $M_{ij}=c(\zb_i,\zb_j)$. The optimization in~\eqref{eq:fixed_support_barycenter} is widely known as a \textit{fixed support barycenter} problem~\citep{agueh2011barycenters,cuturi2014fast}. 

We can further rewrite~\eqref{eq:fixed_support_barycenter} as:
\begin{align*}
&\min_{\boldsymbol{\alpha} \in \Delta^{|\ZZ|} } \sum_{k=1}^K \lambda_k \min_{\gamma_k \in \Gamma(\boldsymbol{\alpha},\boldsymbol{\alpha}^{(k)})}  \langle \gamma_k,M\rangle -\epsilon_k E(\gamma_k) 
\end{align*}
where $\Gamma (\boldsymbol{\alpha},\boldsymbol{\alpha}^{(k)})
=\{\gamma_k \in \mathbb{R}_+^{|\ZZ| \times |\ZZ|} \mid \gamma_k
\mathbf{1} =  \boldsymbol{\alpha},\gamma_k^\top \mathbf{1}=
\boldsymbol{\alpha}^{(k)} \}$ and $E(\gamma_k) =
-\sum_{i=1}^{|\ZZ|}\sum_{j=1}^{|\ZZ|} \gamma_{k,ij}
\log\gamma_{k,ij}$
 and $\langle \gamma_k,M\rangle$ is the sum over $\ZZ \times \ZZ$
as in~\eqref{eq:eOT}. 
Let $\boldsymbol{\gamma}
=(\gamma_1,\ldots,\gamma_K)$, the optimization is equivalent to: 
\begin{align*}
\min_{\boldsymbol{\gamma}} \sum_{k=1}^K \lambda_k \left[\langle \gamma_k,M\rangle -\epsilon_k E(\gamma_k)\right], 
\end{align*}
subject to  $\gamma_k\in  \mathbb{R}_+^{|\ZZ| \times |\ZZ|} $,
$\gamma_k^\top \mathbf{1} = \boldsymbol{\alpha}^{(k)}$  for
$k=1,\ldots,K$, and $\gamma_1\mathbf{1}=\gamma_2 \mathbf{1}=\ldots=
\gamma_K \mathbf{1}$. We can interpret the constraint of
$\boldsymbol{\gamma}$ as the intersection of  two constraint sets
$\mathcal{C}_1=\{\boldsymbol{\gamma} \in (\mathbb{R}_+^{|\ZZ| \times
  |\ZZ|})^K\mid  \gamma_k^\top \mathbf{1} = \boldsymbol{\alpha}^{(k)},
\forall k=1,\ldots,K\}$ and $\mathcal{C}_2 = \{\boldsymbol{\gamma} \in
(\mathbb{R}_+^{|\ZZ| \times |\ZZ|})^K\mid  \exists \boldsymbol{\alpha}
\in \Delta^{|\ZZ|},  \gamma_k \mathbf{1} = \boldsymbol{\alpha},
\forall k=1,\ldots,K\}$.  By defining
$\KL_{\lamb}(\boldsymbol{\gamma},\boldsymbol{\xi}) = \sum_{k=1}^K
\lambda_k \KL(\gamma_k,\xi_k)$ and defining
$\KL(\gamma_k,\xi_k)=\sum_{i=1}^{|\ZZ|} \sum_{j=1}^{|\ZZ|}
\gamma_{k,ij} \left(\log\frac{\gamma_{k,ij}}{\xi_{k,ij}}-1\right)$
(abusing $\KL$ notation), we can write the optimization problem into
the final form: 
\begin{align}
\label{eq:KL_IBP}
     \min_{\boldsymbol{\gamma} \in \mathcal{C}_1\cap \mathcal{C}_2} \KL_{\lamb}(\boldsymbol{\gamma},\boldsymbol{\xi}), 
\end{align}
where $\xi_k = \exp(-M/\epsilon_k)$.

The optimization in~\eqref{eq:KL_IBP} is known to have an unique solution, which can be obtained using iterative Bregman projections~\citep{benamou2015iterative}. In summary, the algorithm iteratively perform two steps:
 (1) $\boldsymbol{\gamma}=\argmin_{\boldsymbol{\gamma} \in \mathcal{C}_1} \KL_{\lamb}(\boldsymbol{\gamma},\boldsymbol{\xi})$, and
   (2) $\boldsymbol{\gamma}=\argmin_{\boldsymbol{\gamma} \in \mathcal{C}_2} \KL_{\lamb}(\boldsymbol{\gamma},\boldsymbol{\xi})$. The above two steps have the following closed-form updates:  for $k=1,\ldots,K$,
 \begin{enumerate}
    \item $\gamma_k = \gamma_k \text{diag} \left(\frac{\boldsymbol{\alpha}^{(k)}}{\gamma_k^\top \mathbf{1}}\right)$,
    \item $\gamma_k = \text{diag}\left(\frac{\boldsymbol{\alpha}}{\gamma_k \mathbf{1}}\right) \gamma_k$ where $\alpha=\prod_{k=1}^K (\gamma_k \mathbf{1})^{\lambda_k}$ ($\prod$ and $(\cdot)^{\lambda_k}$ are entry-wise operators).
\end{enumerate}
As mentioned, the algorithm converges to a unique solution as the
number of iterations tends to infinity. In practice, there are some
 clever implementation choices  such as normalizing
$\boldsymbol{\alpha}$ each iteration, and memory efficient and
parallel implementation (see details in~\citep{benamou2015iterative}).  

Considering all partitions in $\ZZ$ as atoms of distributions is
computationally  impractical,  as the number of possible
partitions is enormous. Therefore, we need to restrict partitions to a
feasible set. For example, we 
 use the much smaller sets of posterior MCMC
samples from $p_k(\zb \mid X^{(k)})$,  or samples from
its variational approximations. In particular, letting
$\zb_1^{(k)},\ldots,\zb_{N_k}^{(k)} \sim p_k(\zb \mid X^{(k)})$
($N_k>0$ small compared to $|\ZZ|$), we can approximate
$p_k(\zb \mid X^{(k)}) \approx \frac{1}{N_k}
\sum_{i=1}^{N_k}\delta_{\zb_i^{(k)}}$ and consider finding the
consensus posterior $\bar{p}(\zb\mid X,\lamb) =  \sum_{k=1}^K
\sum_{i=1}^{N_k} \alpha_{ki} \delta_{\zb_i^{(k)}}$ where
$(\alpha_{11},\ldots,\alpha_{1N_1},\ldots,
\alpha_{K1},\ldots,\alpha_{KN_K})\in \Delta^{K N_k}$. As we do not
need the consensus barycenter to have shared atoms with shard
posteriors, we can further reduce the number of atoms of the consensus
posterior  using the shard posteriors
summarization~\citep{wade2018bayesian,dahl2022search,nguyen2026summarizing,balocchi2025understanding}
or random sampling. For these cases, we can just update
$\boldsymbol{\xi}$ in~\eqref{eq:KL_IBP} i.e., $\xi_k =
\exp(-M^{(k)}/\epsilon_k)$ with $M^{(k)}$ being the ground cost matrix
between the atoms of the consensus posterior and the atoms of the
$k$-th shard posterior. We refer the reader to Figure~\ref{fig:VCI}
for a simple illustration of VCI. In principle, we can also search for
the atoms of the consensus posterior during optimization (known as
\textit{free support barycenter}~\citep{cuturi2014fast}); however,
search on the space of partitions $\ZZ$ is expensive. Therefore, we
leave this investigation to future work.

\section{Experiments}
\label{sec:exp}
We demonstrate VCI in three scenarios: \textbf{scenario 1}
validates VCI as an approximate of model-based clustering in
low dimensional problems. \textbf{Scenario 2} verifies that
VCI can recognize the noisy dimensions in a moderate dimensional problem with
irrelevant featurs. 
And \textbf{scenario 3} is designed to validate VCI as meaningful
inference with high-dimensional data. \ech

\begin{table}[!t]
    \centering
    \caption{Scenario 1. Results on Old Faithful Geyser Dataset.
      The right column reports distances to the ground truth $p_0$.}
    \scalebox{0.9}{
    \begin{tabular}{|l|c|}
    \toprule
         Posteriors& $W_{\text{VoI}}(\cdot,p_0(\zb\mid X))$\\
             \midrule
       $p_1(\zb \mid X^{(1)})$  & 0.3030 \\
       $p_2(\zb \mid X^{(2)})$  & 0.3629\\
           \midrule
       $\frac{1}{K}\sum_{k=1}^K p_k(\zb\mid X^{(k)} )$&0.3142\\
       $\bar{p}(\zb\mid X,\boldsymbol{1/K})$  & 0.2510\\ 
       $\sum_{k=1}^K \lambda^{H}_kp_k(\zb\mid X^{(k)} )$&0.3139\\
       $\bar{p}(\zb\mid X,\boldsymbol{\lambda}^H)$  & 0.2501\\ 
       $\sum_{k=1}^K \lambda^{P}_kp_k(\zb\mid X^{(k)} )$&0.3109\\
       $\bar{p}(\zb\mid X,\boldsymbol{\lambda}^{P})$  & \textbf{0.2465}\\ 
       \bottomrule
    \end{tabular}
    }
    \label{tab:faithful}
    \vspace{-1em}
\end{table}

 \subsection{Scenario 1}   
 For an easy and well-recognized example  
we consider the two-dimensional Old Faithful Geyser
dataset~\citep{azzalini1990look}. We fit conjugate truncated Dirichlet
process mixture of Gaussians models~\citep{ishwaran2001gibbs} to these
datasets.  
The posterior distribution $p_0(\zb\mid X)$ of the random partition based on the full data is treated as the ground truth, represented as an empirical distributions over 1,000 MCMC samples after a burn-in of 9,000 samples. Hyperparameters are kept consistent between the full model and the models for shards. We partition the data into two one-dimensional shards. For each shard, we obtain 1,000 MCMC samples after a burn-in of 9,000 iterations. The consensus posterior is constructed as a discrete distribution supported on the union of the two sets of MCMC samples, with weights determined by~\eqref{eq:fixed_support_barycenter} using $\epsilon_k = 0.05$ (the smallest value that remains numerically stable). For the barycenter weights, we consider three choices: uniform weights, and weights based on the entropy $\lamb^H$, and the proposed weights  $\lamb^P$ ($a=1$ as the default choice), as described in Section~\ref{subsec:proposed_methods}. We report Wasserstein distances, using the variation of information (VoI) ground metric, between the target posterior and the shard posteriors, their mixtures, and the consensus posterior in Table~\ref{tab:faithful}. We observe that the consensus posteriors achieve the smallest distances, outperforming even the best individual shard posterior. The weights $\lamb^P$ provide the lowest distance.

\begin{table}[!t]
    \centering
    \caption{Scenario 2. Results on Noisy Old Faithful Geyser Dataset.
      The last column reports distancde to the first two meaningful
      dimensions. }
    \scalebox{0.9}{
    \begin{tabular}{|l|c|}
    \toprule
        Posterior & $W_{\text{VoI}}(\cdot,p_1(\zb\mid X^{(1)}))$\\
         \midrule
       $p_0(\zb\mid X)$  & 0.9464 \\
       $p_1(\zb \mid X^{(1)})$  & 0\\
       $p_{k}(\zb\mid X^{(k)}),$ $k=2,\ldots,10$ & [0.8847, 3.4301]\\
       \midrule
       $\frac{1}{K}\sum_{k=1}^K p_k(\zb\mid X^{(k)} )$&1.9794\\
       $\bar{p}(\zb\mid X,\boldsymbol{1/K})$  & 0.8850\\ 
       $\sum_{k=1}^K \lambda^{H}_kp_k(\zb\mid X^{(k)} )$&3.2301\\
       $\bar{p}(\zb\mid X,\boldsymbol{\lambda}^H)$  & 0.8987\\ 
       $\sum_{k=1}^K \lambda^{P}_kp_k(\zb \mid X^{(k)} )$&0.2154\\
       $\bar{p}(\zb\mid X,\boldsymbol{\lambda}^{P})$  & \textbf{0.0031}\\ 
       \bottomrule
    \end{tabular}
    }
    \label{tab:faithful_noise}
    \vspace{-1.5em}
\end{table}

\subsection{Scenario 2}

We add 18 dimensions to the Old Faithful Geyser dataset. The first 10
dimensions are from sampling $n$ observations from a  Gaussian with
mean $(3,70,\ldots,3,70)^\top$ and covariance
$\text{diag}([4,36,\ldots,4,36])$ while the last 8 dimensions are from
sampling $n$ observations from a  Gaussian with mean
$(1,10,\ldots,1,10)^\top$ and covariance
$\text{diag}([1,4,\ldots,1,4])$.  We then repeat the same procedure as
in \textbf{scenario 1} but with 10 shards of 2 dimensions. For each
shard, we  save 100 posterior MCMC samples after a burn-in of
9,900. 
Table~\ref{tab:faithful_noise} reports Wasserstein distances to
the clean posterior (under the first shard).  For shards
 including noisy dimensions 
the distances are from 0.88 to 3.43. The consensus barycenters
mitigates the  contamination with the  noisy dimensions, as
desired. 
For the choice of barycenter weights, the
weights $\lamb^P$ ($a=10$ to penalize  large $K$) 
result in the lowest distance.




\subsection{Scenario 3}
We consider high-dimensional single-cell data with 25,348
dimensions/genes~\citep{10xGenomics2024}. After pre-processing, we
obtain 800 cells from 10 cell-types
annotated by known cell markers (observed pseudo labels).
 For model-based inference on random partitions we fit 
the following model: 
\begin{align*}
X_i \mid \boldsymbol{\theta}_i 
&\sim \prod_{d=1}^p \mathrm{Poisson}(N_i \theta_{i d}), \\
\boldsymbol{\theta}_i \mid G 
&\sim G, 
G 
\sim \mathrm{DP}\left(\alpha, \prod_{d=1}^p \mathrm{Gamma}(a,b)\right),
\end{align*}
where $N_i = \sum_{d=1}^p X_{id}$ is the sequencing depth. We consider
$K=100$ shards with approximately equal dimensions. We fit the model
on the full data and on each shard,  saving 100 MCMC samples after
a burn-in of 900. 
For consensus posteriors we use the last 10 MCMC
samples per shard to construct
 empirical shard posteriors (over partitions) with 
1,000 atoms. For the barycenters, we
use $\epsilon_k = 0.05$ for the entropic regularization. In
Table~\ref{tab:single_cell}, we report expected VoI distances
 relative to the 
observed pseudo labels. For the full posterior, the expected distance
is 2.3026. For shards, the expected distances are from 1.1897 to
2.5237. All consensus posteriors achieve lower expected distances than
the full posterior. Regarding the choice of barycenter weights, those
based on entropy yield the smallest distances, as all shards produce
relatively few clusters compared to the pseudo labels. We also observe
that smaller shards tend to generate more clusters; however, this does
not increase the expected VoI distance to the pseudo truth. It is
important to note that the labels are pseudo, so they do not fully
capture the true quality of the methods.

\begin{table}[!t]
    \centering
    \caption{Scenario 3. Results on Single Cell Dataset.
      The last column reports distance to the pseudo truth.}
    \scalebox{0.9}{
    \begin{tabular}{|l|c|}
    \toprule
        Posteriors & $\mathbb{E}[\text{VoI}(\cdot,\zb^\star)]$\\
         \midrule
       $p(\zb\mid X)$  & 2.3026 \\
       $p_{k}(\zb\mid X),$ $k=1,\ldots,100$&[1.1897, 2.5237] \\
       \midrule
       $\frac{1}{K}\sum_{k=1}^K p_k(\zb \mid X^{(k)} )$&1.7901\\
       $\bar{p}(\zb\mid X,\boldsymbol{1/K})$  & 1.7756\\ 
       $\sum_{k=1}^K \lambda^{H}_kp_k(\zb \mid X^{(k)} )$&1.6036\\
       $\bar{p}(\zb\mid X,\boldsymbol{\lambda}^{H})$  & \textbf{1.5496}\\ 
       $\sum_{k=1}^K \lambda^{P}_kp_k(\zb \mid X^{(k)} )$&1.7375\\
       $\bar{p}(\zb\mid X,\boldsymbol{\lambda}^P)$  & 1.7194\\ 
       \bottomrule
    \end{tabular}
    }
    \vspace{-1.5em}
    \label{tab:single_cell}
\end{table}

\section{Conclusion}
\label{sec:conclusion}

VCI provides a principled and scalable framework for high-dimensional
Bayesian clustering. By operating on vertical shards and leveraging an
entropic-regularized Wasserstein barycenter, it mitigates the curse of
dimensionality while retaining interpretability through a generalized
Bayes perspective. The framework favors shards with desired properties through
controlling the barycenter weights. Empirical results demonstrate that
VCI achieves 
accurate approximations in low dimensional settings and
performs favorably in high-dimensional scenarios with many irrelevant
features. Future work may explore alternative regularized Wasserstein
barycenter formulations~\citep{bresch2026interpolating}, which could
correspond to variational inference under different probabilistic
discrepancies beyond the KL
divergence~\citep{knoblauch2022optimization}. Another direction is to
iteratively update both variational and consensus posteriors
in~\eqref{eq:prop_equation}, rather than fixing the variational
distributions to be the shard posteriors. Developing
principled approaches for constructing problem-specific barycenter
weights remains an important challenge. Extending VCI beyond random
partitions to more general dimension-independent parameterizations is
another promising direction. Finally, improving the computational efficiency
of barycenter estimation, e.g., through sliced optimal
transport~\citep{bonneel2015sliced,nguyen2025introduction}, is another
important direction for future work.

 \begin{appendix}
 
 \subsection*{Appendix A. Proof of Proposition 2.1}
First, we utilize the fact that the entropy of  \textit{discrete} random variables is non-negative to obtain a lower bound of the joint entropy $H_q(\zb,\zb^{(1):(K)})$:
\begin{multline} \label{eq:entropy_inequality} 
    H_q(\zb,\zb^{(1):(K)}) = H(\zb) +H_q(\zb^{(1):(K)}\mid \zb) \\
    =  H_q(\zb) +\frac{1}{K} \sum_{k=1}^K [H_q(\zb^{(k)}\mid \zb) 
    + H_q(\zb^{-(k)}\mid \zb,\zb^{(k)})]\\
    \geq H(\zb) 
    +\frac{1}{K} \sum_{k=1}^K H_q(\zb^{(k)}\mid \zb)
    =\frac{1}{K}\sum_{k=1}^K H_q(\zb,\zb^{(k)}),  
\end{multline}
where $H_q(\zb^{(1):(K)}\mid \zb) = -\mathbb{E}_{(\zb,\zb^{(1):(K)})\sim q}[ \log q(\zb^{(1):(K)}\mid \zb )]$ is the conditional entropy. We then define the upper bound:
\begin{align*}
    \overline{\mathcal{L}}(q) &= \mathbb{E}_{(\zb,\zb^{(1):(K)})\sim q}[-\log p(\zb,\zb^{(1):(K)})] \nonumber \\
    &-\frac{1}{K}\sum_{k=1}^K H_q(\zb,\zb^{(k)}) \geq \mathcal{L}(q).
\end{align*}
We further have $     \log p(\zb,\zb^{(1):(K)},X)=$
\begin{multline}
  = \log\left[ p(\zb) \prod_{k=1}^K p(\zb^{(k)}\mid \zb) p_k (X^{(k)} \mid \zb^{(k)}) \right]\\ 
     = \sum_{k=1}^K -\zeta_k c(\zb^{(k)},\zb) + \log  p_k (X^{(k)}
     \mid \zb^{(k)}) -C,
     \nonumber
\end{multline}
for a constant $C=\sum_{\zb\in\ZZ} \prod_{k=1}^K C_k(\zb) >0$. Let
$q_{0,k}(\zb,\zb^{(k)}) \in \Pi(q_0,q_k)$ be the marginal of the joint
$q(\zb,\zb^{(1):(K)})$ at $(\zb,\zb^{(k)})$, and
$\Pi^*(q_0,q_1,\ldots,q_K)$ be the set of star couplings such that
$q(\zb, \zb^{(1)}, \dots, \zb^{(K)}) = q_0(\zb) \prod_{k=1}^K
q(\zb^{(k)} \mid \zb)$ for any $q\in \Pi^*$, we have:
{\small
\begin{align}
      &\min_{q \in \Pi(q_0,q_1,\ldots,q_K)}{\mathcal{L}}(q) \leq  \min_{q \in \Pi(q_0,q_1,\ldots,q_K)}\overline{\mathcal{L}}(q) \nonumber \\
   & = \min_{q \in \Pi(q_0,q_1,\ldots,q_K)} \mathbb{E}_{(\zb,\zb^{(1):(K)})\sim q}\left[\sum_{k=1}^K \zeta_k c(\zb^{(k)},\zb) \right. \nonumber\\
    &\left.- \log  p_k (X^{(k)} \mid \zb^{(k)})+\frac{1}{K}\log q(\zb,\zb^{(k)})\right]+C \nonumber \\
&\leq \min_{q \in \Pi^*(q_0,q_1,\ldots,q_K)} \mathbb{E}_{(\zb,\zb^{(1):(K)})\sim q}\left[\sum_{k=1}^K \zeta_k c(\zb^{(k)},\zb) \right. \label{eq:star_inequality} \\
   & \left.- \log  p_k (X^{(k)} \mid \zb^{(k)})+\frac{1}{K}\log q(\zb,\zb^{(k)})\right] +C\nonumber \\
&= \min_{q \in \Pi^*(q_0,q_1,\ldots,q_K)} \sum_{k=1}^K\mathbb{E}_{(\zb,\zb^{(1):(K)})\sim q}\left[ \zeta_k c(\zb^{(k)},\zb) \right.  \nonumber\\
   & \left.- \log  p_k (X^{(k)} \mid \zb^{(k)})+\frac{1}{K}\log q(\zb,\zb^{(k)})\right]+C \nonumber \\
    &=  \sum_{k=1}^K \min_{q_{0,k}}\mathbb{E}_{(\zb,\zb^{(k)})\sim
      q_{0,k}}\left[  \zeta_k c(\zb^{(k)},\zb)
      - \log  p_k (X^{(k)} \mid \zb^{(k)})+
       \right. \nonumber\\ &\left.
      + \frac{1}{K}\log q(\zb,\zb^{(k)})\right] +C\nonumber \\
    &=  \sum_{k=1}^K \zeta_k \left\{ \min_{q_{0,k}}\mathbb{E}_{(\zb,\zb^{(k)})\sim q_{0,k}}\left[  c(\zb^{(k)},\zb) \right]\right. 
    \left.- \frac{1}{ K \zeta_k}H_{q_{0,k}}(\zb,\zb^{(k)}) \right\}  \nonumber \\
   & - \sum_{k=1}^K\mathbb{E}_{ \zb^{(k)}\sim q_k}\left[\log  p_k (X^{(k)} \mid \zb^{(k)})\right]+C \nonumber \\
    &= \sum_{k=1}^K \zeta_k W_{c,\frac{1}{ K \zeta_k}}(q_0,q_k)  \\&
    - \sum_{k=1}^K\mathbb{E}_{ \zb^{(k)}\sim q_k}\left[\log  p_k (X^{(k)} \mid \zb^{(k)})\right] +C, \nonumber
  \end{align}
  }
which completes the proof.

\end{appendix}

\bibliographystyle{imsart-number} 
\bibliography{bib}       





\end{document}